\documentclass[5p]{elsarticle}

\usepackage{graphicx}
\usepackage{amsmath}
\usepackage{amssymb}
\usepackage{amsfonts}
\usepackage{mathtools}
\usepackage{hyperref}
\usepackage{xcolor}
\usepackage{soul}
\usepackage{bm}
\usepackage{esvect}

\journal{Astronomy and Computing}
\biboptions{authoryear}

\begin{document}

\begin{frontmatter}

\title{Cosmic Voids in GAN-Generated Maps of Large-Scale Structure}

\author[mymainaddress]{Olivia Curtis\corref{mycorrespondingauthor}}
\cortext[mycorrespondingauthor]{Corresponding author}
\ead{amcurtis@bu.edu}
\author[mymainaddress]{Tereasa G. Brainerd}
\author[mysecondaryaddress]{Anthony Hernandez}

\address[mymainaddress]{Department of Astronomy \& Institute for Astrophysical Research, Boston University, Boston, MA 02215, USA}
\address[mysecondaryaddress]{Department of Computer Science and Engineering, University of South Florida, Tampa, FL 33620, USA}

\begin{abstract}

A Generative Adversarial Network (GAN) was used to investigate the
statistics and properties of voids in a $\Lambda$CDM universe.  
The total 
number of voids and the distribution of void sizes
is similar in both sets of images and, within the formal error bars,
the mean void properties are
consistent with each other.
However, the generated images yield somewhat fewer
small voids than do the simulated images. 
In addition, the generated images yield far fewer voids with 
central density contrast $\sim -1$.  Because the generated images yield
fewer of the emptiest voids, the distribution of the mean interior density
contrast is systematically higher for the generated voids than it is
for the simulated voids.
The mean radial underdensity 
profiles of the largest voids are similar in both 
sets of images, but systematic differences are apparent.  On small 
scales ($r < 0.5 r_v$), the underdensity profiles of the voids in 
the generated images exceed those of the voids in the simulated
images.  On large scales ($r >  0.5 r_v$), the underdensity
profiles of the voids in the simulated images exceed those of
the voids in the generated images.
The discrepancies between the void properties in the two sets of
images are attributable to the GAN 
struggling to capture absolute patterns in the data. In particular,
the GAN produces too few pixels with density contrasts $\sim -1$
and too many pixels with density contrasts in the range $\sim -0.88$
to $\sim -0.63$.  

\end{abstract}


\end{frontmatter}

\section{Introduction} \label{sec:intro}

Galaxies are distributed within an interconnected, large-scale network of walls and filaments that stretch for hundreds of megaparsecs. Between these structures lie vast, underdense regions of space known as voids. Voids can reach up to $100 h^{-1}$Mpc in diameter (see, e.g., \citealt{1991ARA&A..29..499G} and references therein), and they have the potential to serve as excellent laboratories for testing the popular $\Lambda$ Cold Dark Matter ($\Lambda$CDM) model of structure formation. Due to their underdense nature, voids are dominated by dark energy 
and they are only weakly influenced by the non-linear effects of gravity (see, e.g., \citealt{goldberg2004}). Because of this, the shapes and distributions of voids provide
tests of modified theories of gravity, as well as constraints on the dark energy equation of state, inflationary models, the sum of the neutrino masses, and the expansion rate and geometry of the universe (see, e.g., \citealt{li2012haloes}, \citealt{sutter2012},
\citealt{clampitt2013voids}, \citealt{cai2015testing}, \citealt{mao2017cosmic},
\citealt{falck2018using}, \citealt{sahlen2019cluster}). Furthermore, the physical properties of void galaxies provide critical insight into the history of galaxy formation and evolution (see, e.g., \citealt{hoyle2005}, \citealt{kreckel2012}, \citealt{douglass2017}, \citealt{tavasoli2021})
.

The total number of voids in existing void catalogs ranges from
$\sim 10^2$ (e.g., \citealt{hoylevoidsin2dfs}, \citealt{sanchez2016cosmic}) to
$\sim 10^3$ to $\sim 6\times 10^3$ (e.g., \citealt{mao2017cosmic}, \citealt{Aubert2020}, \citealt{Hamaus2017})  Near-future surveys, such as those that will be carried out by the Vera C.\ Rubin Observatory 
(\citealt{LSST}),
the Nancy Grace Roman telescope (\citealt{Spergel2015}), the Euclid satellite (\citealt{Euclid}),
the SPHEREx mission (\citealt{Dore2018}), the DESI experiment (\citealt{DESI}), and the Prime Focus
Spectrograph (\citealt{Tamura2016}) on the Subaru telescope, are expected to each yield 
void catalogs containing at least $\sim 10^5$ voids.
This dramatic increase in data 
should lead to significant improvements in our understanding of the properties of voids and void galaxies in the observed universe. 

From the standpoint of large-scale structure theory, gigaparsec-scale simulations are required
in order to determine whether $\Lambda$CDM is able to successfully reproduce the largest structures in the universe.  While Gpc-scale N-body mock catalogs do exist (e.g., \citealt{kim2011new}, \citealt{Indra}), the largest voids are so rare that only a few dozen of these objects are found within a single Gpc-scale simulation. N-body simulations of this
size are computationally expensive to run, making it challenging to use these types of mock catalogs to fully investigate the statistics of the largest voids in $\Lambda$CDM universes.
In order to make progress on the theoretical properties of the largest voids, it would be helpful to be able to use a modeling technique that is capable of quickly producing independent, novel catalogs of large-scale structure.   One technique that may prove advantageous for such applications is the use of 
Generative Adversarial Networks (GANs).

GANs have already begun to demonstrate their usefulness for astrophysical applications. For example, \cite{rodriguez2018fast}, \cite{curtis2020fast}, \cite{feder2020nonlinear}, and \cite{kodi2020super} used GANs to produce novel images of the large-scale structure of the universe. In addition, \cite{mustafa2019cosmogan} used GANs to generate novel weak lensing convergence maps.  Further, \citealt{schawinski2017generative}, \cite{zingales2018exogan}, and \citealt{gan2021seeinggan} trained GANs to map observational data to either super-resolved data or denoised data, allowing computationally expensive Bayesian analyses to be bypassed in some cases.


In order for GANs to develop into computationally inexpensive alternatives to enormous, Gpc-scale N-body simulations, their efficacy in reproducing large-scale structure statistics must first be tested on smaller, more typical N-body simulations.
Generally, the first assessment of GAN performance is a visual examination of the model's outputs (i.e., `by-eye' inspection of the outputs).  This is then followed by comparisons of lower-order and higher-order statistics for a generated sample and the training set. In the context of studies that use GANs to reproduce large-scale structure, lower-order statistics would include matter power spectra and/or mass density histograms (see, e.g., \citealt{rodriguez2018fast}, \citealt{feder2020nonlinear}).  Higher-order statistics would include Minkowski functionals such as those used by \cite{mustafa2019cosmogan} to assess their weak lensing convergence maps.

In this paper
we explore the use of void statistics as a method of assessing the ability of a GAN to produce large-scale structure density maps.
We examine the statistics and properties of voids in a $\Lambda$CDM universe, and we
compare results obtained from N-body 
simulations to results obtained from a suitably trained
GAN. Throughout, we search for voids in 2D slices of the large-scale structure to be consistent with modern observational void surveys (see, e.g., \citealt{sanchez2016cosmic}, \citealt{mao2017cosmic}). The paper is organized as follows.  In \S2, we discuss our computational methods.
These include the N-body 
simulations that were used to train the GAN and the details of the particular GAN that we adopted.  We also
discuss the network architecture and optimization of the network weights.  In \S3
we discuss the underdensity probability function, the specific algorithm we used to identify voids, and various void properties.
A discussion of our results is presented in \S4. Throughout, we adopt the following cosmological
parameters: $h = 0.7$, $\Omega_\Lambda = 0.7$,
$\Omega_{m0} = 0.3$, and $\sigma_8 = 0.9$.

\section{Methods} \label{sec:methods}

\subsection{N-body Simulations} \label{sec:nbody}
Before a GAN can be used to generate density maps 
of $\Lambda$CDM universes, 
it first must be taught the
properties that are expected for the large-scale structure.  
Teaching the GAN these expected properties is a process known
as training.  Training the GAN requires a set 
of independent N-body simulations, the results of which
allow the GAN to learn the expected properties of the 
large-scale structure and then to extrapolate 
from its training in order to produce new images.  
Here, the training set consisted of ten $\Lambda$CDM simulations. Each simulation adopted a 
cubical box with periodic boundary conditions, a box length of $L = 512 h^{-1}$Mpc, and
a total of $512^3$ dark matter particles of mass $m_p = 8.3 \times 10^{10}h^{-1}M_\odot$.
A unique set of initial conditions for each simulation was generated using N-GenIC\footnote{\url{https://www.h-its.org/2014/11/05/ngenic-code/}} \citep{2005Natur} and the  well-tested Gadget-2\footnote{\url{https://wwwmpa.mpa-garching.mpg.de/gadget/}} code \citep{springel2005cosmological} 
was used to evolve
the simulations from redshift $z=50$ to the present epoch. 
All simulations were run on the Shared Computing Cluster at the Massachusetts Green 
High-Performance Computing Center.

To create a training set for the GAN, 15,000 2D images were extracted from the N-body simulations using the method adopted by \cite{rodriguez2018fast}. First, each simulation was divided each into 1,000 slices along each of the three primary axes. Each slice was then pixellated using a nearest neighbor mass assignment  scheme and 500 non-consecutive slices along each axis were selected. Lastly, each image was smoothed with a Gaussian filter with a width of 1 pixel. 

Note: while slices that are nearby to one another in a given simulation may be spatially correlated (i.e., sequential slices are separated by distances of $\sim 0.5h^{-1}$Mpc), this is not expected to influence the training of the GAN in any significant way.  
During each training epoch, the order of
the 15,000 slices is randomized and batches of 16 sequential images from the randomized slices are then fed to the network prior to back propagation (see Table~1).  Because of this, it is unlikely that any of the 16 images in a given batch will have any remaining spatial correlation.

Before being passed to the GAN for training, the density maps from the N-body simulations must be normalized.  This is 
necessary because normalizing reduces the cost of transforming the latent space distribution into the space
expected of the true data distribution. Following
\cite{rodriguez2018fast}, the density
maps from the N-body simulations
were normalized using
\begin{equation} \label{eq:norm}
    s(x) = \frac{2x}{x+a} ~~.
\end{equation} 
Here, $a$ is an integer that controls the median pixel value of the normalized map, and the properties 
of $s(x)$ resemble those of a logarithmic function. 

\begin{figure*}
    \centering
    \includegraphics[scale=1.2]{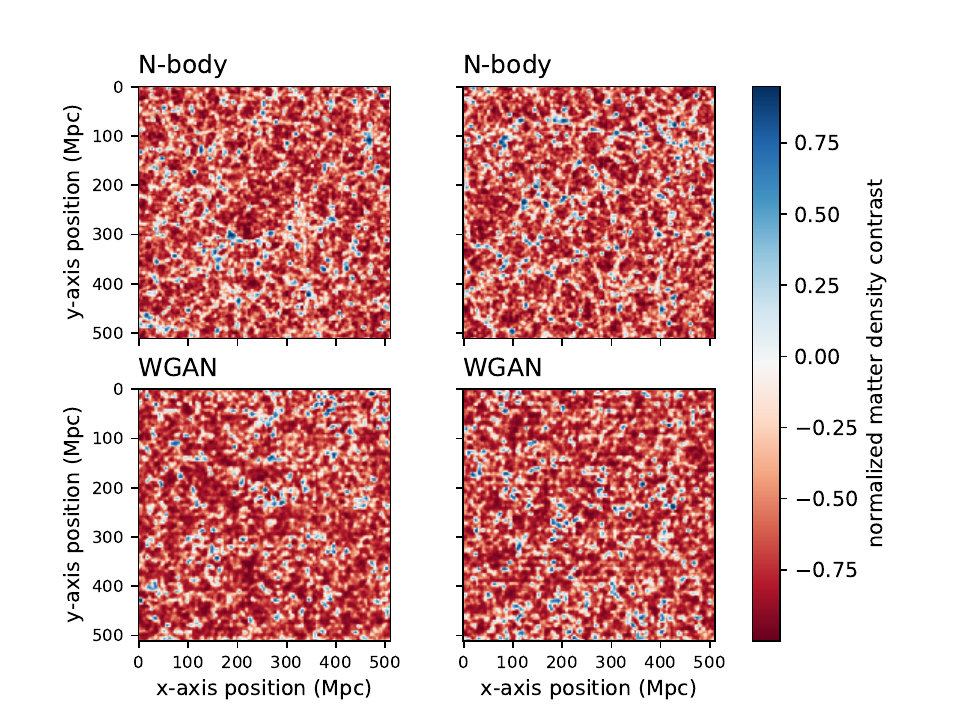}
    \caption{Examples of normalized matter density maps (i.e., `images') at 
    redshift $z = 0$. Each image has a sidelength of
    $512 h^{-1}$~Mpc and a thickness of $0.512 h^{-1}$Mpc. {\it Top:} Images obtained
    from N-body simulations.  {\it Bottom:} Images obtained from a trained Wasserstein GAN (WGAN). 
    Here the local matter density has been normalized using Equation \ref{eq:norm}
    (see text). 
    }
    \label{fig:ex_density_map}
\end{figure*}

\cite{feder2020nonlinear} investigated the effect of changing the value of the scaling parameter in Equation~\ref{eq:norm} ($a$ in our notation, $\kappa$ in \citealt{feder2020nonlinear}).  From this, \cite{feder2020nonlinear} found that larger values of the scaling parameter better preserved the high-density features in their generated density maps.  On the small scales associated with galaxies, this gave rise to an improved agreement between the matter power spectra obtained from the generated images and the matter power spectra obtained from the N-body images. However, on the large scales that are relevant to cosmic voids, \cite{feder2020nonlinear} found that changes in the scale parameter had little affect on power spectrum from the generated images.  Therefore, for consistency with \cite{rodriguez2018fast} we adopt $a = 4$ throughout this paper.
Examples of normalized 2D mass density maps, obtained from the
N-body simulations, are shown in the top panels of Figure \ref{fig:ex_density_map}.

\subsection{Generative Adversarial Networks} \label{sec:gans}

A GAN is a game between two deep convolutional neural networks: a discriminator network ($D$) and a 
generator network ($G$). The discriminator, $D : (\vv{\bm{x}}, \theta_{D}) \rightarrow [0;1]$, attempts to 
label a sample, $\vv{\bm{x}}$, as being either `real' or `fake.' Here, `real' means $\vv{\bm{x}}$ is 
drawn from the true data distribution ($p_{data}$) and `fake' means $\vv{\bm{x}}$ is a generated image drawn 
from the set of generated data ($p_{gen}$). The variable $\theta_{D}$ represents the hyper-parameters that 
characterize $D$. The generator network, $G : (\vv{\bm{z}}, \theta_{G}) \rightarrow \vv{\bm{x}}$, attempts 
to map a random variable, $\vv{\bm{z}}$, to a sample $\vv{\bm{x}}$ that appears to be drawn from $p_{data}$. 
The random variable
is drawn from a latent prior space, $p_{prior}(\vv{\bm{z}})$, which is generally 
taken to be Gaussian-distributed.

The discriminator network is trained to maximize the probability of correctly labelling samples that 
are drawn from the training set and samples that are drawn from the 
generator network. Meanwhile, the generator network is trained to minimize $\log_{10}(1 - D(G(\vv{\bm{z}})))$; 
i.e., the probability of the generator network producing a sample that the discriminator network mislabels 
as `real'. Formally, $D$ and $G$ play what is known as the two-player minimax game: 
\begin{equation} \label{eq:gan}
    \begin{array}{c c}
         min & max\\
         \textsuperscript{G} & \textsuperscript{D}
    \end{array} [L(G_{\theta_G},D_{\theta_D})]
\end{equation}
with loss function
\begin{multline} \label{eq:v}
    L(G_{\theta_G},D_{\theta_D}) := \mathbb{E}_{\vv{\bm{x}}\sim p_{data}(\vv{\bm{x}})}[\log_{e} D_{\theta_{D}}(\vv{\bm{x}})] + \\ \mathbb{E}_{\vv{\bm{z}}\sim p_{prior}(\vv{\bm{z}})}[
    \log_{e} (1-D_{\theta_D}(G_{\theta_G}(\vv{\bm{z}})))] ~,
\end{multline}
where $\mathbb{E}$ is the expectation function. (Note:
Equation \ref{eq:v} reduces to the Jensen-Shannon divergence between $p_{data}$ and $p_{gen}$; see
\citealt{goodfellow2014generative}.)

\subsubsection{Wasserstein GANs}

For our work, we adopted a particular type of GAN known as a Wasserstein GAN. The motivation
for this choice stems from the phenomenon of
mode collapse. Mode collapse is said to have occurred in a generative network 
when the generator produces only one particular output, regardless of its input. 
For example, suppose a GAN is being trained to generate pictures of rooms that would typically be found in a house (see, e.g., \citealt{gulrajani2017}). Mode collapse occurs when the GAN only generates one image of a kitchen as its sole output.


Wasserstein GANs mitigate the problem of mode collapse by adopting what is known as an `Earth mover's distance' loss function (see, e.g., \citealt{WGAN}). When applied to GANs, the Earth mover's distance allows the generator to find a more stable
way of transforming $p_{gen}$ into $p_{data}$. In return, the discriminator network now scores the `realness' and `fakeness' of an image (i.e., it scores
how well $p_{gen}$ resembles $p_{data}$). For this reason, \cite{WGAN} refer to the discriminator network as a `critic network', a terminology
that we also adopt. WGANs address the problem of mode collapse by not 
rewarding the generator network for its ability to generate a single result that consistently deceives the discriminator network, but by instead scoring the `realness' and `fakeness' of an image.

Additional discussion of Wasserstein GANs and the
Earth Mover's distance is given in \ref{sec:wgan appendix} and references therein.

\subsubsection{Adam Optimizer} \label{sec:Adam}

In order to carry out back propagation, we use a standard 
Adaptive Moment Optimizer known as \textit{Adam} (see \citealt{ADAM}). 
\textit{Adam} is a stochastic gradient descent algorithm that 
updates the network weights using running averages of the first and second moments of the gradients of the 
loss function.  The \textit{Adam} optimizer is a popular choice for performing stochastic gradient descent 
because it is computationally efficient and has low memory requirements.

For given a set of parameters, $w^{(t)}$, and a loss function, $L^{(t)}$, at training iteration, $t$, 
the weights at the $t+1$ training iteration are found by first calculating the biased first and second moments using Equations \ref{eq:biased_1_moment} and \ref{eq:biased_2_moment}, respectively:
\begin{equation}\label{eq:biased_1_moment}
  m_{w}^{t+1} \leftarrow \beta_{1} m_{w}^{t} + (1-\beta_{1})\nabla_{w} L^{t}  
\end{equation}
\begin{equation} \label{eq:biased_2_moment}
    v_{w}^{t+1} \leftarrow \beta_{2}v_{w}^{t} + (1-\beta_{2})(\nabla_{w}L^{t})^{2} .
\end{equation}
Here, $\beta_{1}$, $\beta_{2}$ $\in$ [0,1) are the exponential decay rates for the moment estimates, $m_{w}^{(t)}$ are the biased first moment estimates of the weights at step t, $v_{w}^{(t)}$ are the biased second moment estimates, and $\nabla_{w}$ is the gradient operator with respect to the network parameters.

Typical values for $\beta_{1}$ and $\beta_{2}$ are 0.9 and 0.999, so if $m^0_w$ and $v^0_w$ are initialized to zero, 
a significant bias towards small values in future updates will occur. This bias can be eliminated (making the moments less sensitive to $\beta_1$ and $\beta_2$ at early timesteps) by normalizing the $\beta$'s such that $\sum_{i=0}^{t} \beta^i = \frac{1-\beta^t}{1-\beta}$. 
This normalization can then be removed 
by dividing the first and second moments by $1-\beta^{t+1}$ in Equations \ref{eq:unbiased_1_moment} and \ref{eq:unbiased_2_moment} below:
\begin{equation}\label{eq:unbiased_1_moment}
     \hat{m}_{w} = \frac{m_{w}^{t+1}}{1-\beta_{1}^{t+1}}
\end{equation}
\begin{equation}\label{eq:unbiased_2_moment}
    \hat{v}_{w} = \frac{v_{w}^{t+1}}{1-\beta_{2}^{t+1}} ~.
\end{equation}
Finally, the weights at step $t+1$ are calculated from the unbiased moments using
\begin{equation}\label{eq:weight_update}
    w^{t+1} \leftarrow w^{t} - \alpha \frac{\hat{m}_{w}}{\sqrt{\hat{v}_{w}} + \epsilon} ,
\end{equation}
where $\alpha$ is the algorithm's learning rate and $\epsilon$ is a small number that is used to prevent division by zero.
Table \ref{tab:hyperparams} lists the values
of the hyperparameters that we adopted for our work.

\begin{table*}
    \centering
    \caption{Hyperparmeters used in the \textit{Adam} optimizer, as well as other important network parameters (see Table \ref{tab:architecture})}
    \begin{tabular}{l l l}
    \hline Hyperparameter & Value & Description \\ \hline
    Batch Size & 16 & Number of training samples fed into the network before back propagation\\
    \textbf{z} dimension & 200 & Dimension of the random Gaussian vector\\
    $\alpha_{D}$ & 1e-6 & Learning rate for the discriminator's (D) \textit{Adam} optimizer \\
    $\beta_{1,D}$ & 0.5 & First exponential decay rate for D's \textit{Adam} optimizer \\
    $\beta_{2,D}$ & 0.999 & Second exponential decay rate for D's \textit{Adam} optimizer \\
    $\alpha_{G}$ & 5e-5 & Learning rate for the generator's (G) \textit{Adam} optimizer \\
    $\beta_{1,G}$ & 0.5 & First exponential decay rate for G's \textit{Adam} optimizer \\
    $\beta_{2,G}$ & 0.999 & Second exponential decay rate for G's \textit{Adam} optimizer \\
    $\alpha$ & 0.8 & Momentum used in batch normalization \\
    $n_{disc}$ & 5 & Number of times D is updated for every G update \\
    c & 0.01 & Weight clipping parameter used to uphold Lipschitz continuity in WGAN architectures \\
    $\alpha_{LeakyReLU}$ & 0.2 & Negative slope coefficient for LeakyReLU activation functions \\
    \hline
    \end{tabular}
    \label{tab:hyperparams}
\end{table*}


\subsubsection{Network Architecture}

Throughout, we use a standard WGAN architecture with deep convolutional layers.  Details of 
the architecture are summarized in 
Table~\ref{tab:architecture}. The critic takes an image of size 
$256 \times 256$ pixels as its input. Four 2D convolutional layers then down-sample the image. The output of the 
critic is a score in the range $(-1,1)$, which describes how `real' ($-1$) 
or `fake' ($+1$) the critic thinks the image is.
For the convolutional layers, a $4 \times 4$ pixel kernel with a stride of size $(2,2)$ is used.
In addition, the inputs of the network are batch normalized before each mini-batch update.
The first four convolutional layers have a Leaky Rectified Linear Unit (Leaky-ReLU) activation function, and the last layer has a linear activation function. The first hidden layer down-samples the image into a tensor of shape $128 \times 128 \times 64$, where the last dimension is the number of channels. Each consecutive hidden layer then doubles the number of channels to 128, 256, and 512 respectively.

\begin{table*}
    \centering
    \caption{WGAN network architecture used in this paper}
    \begin{tabular}{l l l l}
    \hline Layer & Operation & Output & Dimension \\ \hline
    Critic & & & \\
    \textbf{X} & & & 256 x 256 x 1 \\
    $D_0$ & Conv2D & LeakyReLU-BatchNorm & 128 x 128 x 64 \\
    $D_1$ & Conv2D & LeakyReLU-BatchNorm & 64 x 64 x 128 \\
    $D_2$ & Conv2D & LeakyReLU-BatchNorm & 32 x 32 x 256 \\
    $D_3$ & Conv2D & LeakyReLU-BatchNorm & 16 x 16 x 512 \\
    $D_4$ & Linear & Linear & 1 \\
    Generator & & & \\
    \textbf{Z} & & & 200 x 1\\
    $G_0$ & Linear & ReLU-BatchNorm & 16 x 16 x 512 \\
    $G_1$ & Conv2D Transpose & ReLU-BatchNorm & 32 x 32 x 256 \\
    $G_2$ & Conv2D Transpose & ReLU-BatchNorm & 64 x 64 x 128 \\
    $G_3$ & Conv2D Transpose & ReLU-BatchNorm & 128 x 128 x 64 \\
    $G_4$ & Conv2D Transpose & Tanh & 256 x 256 x 1 \\
    \hline
    \end{tabular}
    \label{tab:architecture}
\end{table*}

The generator takes a random 1D Gaussian vector of length 200 as its input. The first hidden layer is a linear layer that reshapes the latent space vector into a tensor of shape 
$16 \times 16 \times 512$ (i.e., 512 low resolution images of size $16 \times 16$ pixels). Four 2D convolutional transpose layers up-scale the image while reducing the number of channels. The first four layers use a ReLU activation function and have their inputs batch normalized before each mini-batch update. The output of the generator is an image of size 
$256 \times 256$ pixels where each pixel is cast into the range $(-1,1)$ using a hyperbolic tangent function. To initialize the weights of the critic and generator, a Glorot weight initialization was used. This initializes each weight as a random number, pulled from a normal distribution that is centered on 0 and bounded by the inverse square root of the number of inputs to that node (see, e.g., \citealt{glorot2010understanding}). 

The networks were implemented using Python's TensorFlow\footnote{\url{https://www.tensorflow.org/}} package (e.g., \citealt{python}; \citealt{abadi2016tensorflow}) and were trained on two NVIDIA Tesla P100 graphics processing units with compute capability 6.0. The critic and generator networks had 2.89 million and 16.00 million trainable parameters, respectively. Training was carried out with Python version 3.7.7, TensorFlow version 1.15.0, CUDA version 10.0, and NVIDIA cuDNN version 7.6, and took approximately four hours for 20 training epochs. The implementation of the WGAN described in this paper is available on the first author's GitHub page.\footnote{\url{https://github.com/o-curtis/}}

\section{Results} \label{sec:results}


The WGAN was trained for 20 epochs and the network parameters were saved after each training epoch. After 20 epochs, the outputs of each saved network were analyzed to determine which training epoch produced matter power spectra that were most similar to the training set. In the case of our WGAN, we found this to occur after four training epochs.  This network was then used to generate a set of 15,000 density maps 
(i.e., `images') which were used for the analyses we present below.  Two of these generated images are shown in the bottom panels of
Figure \ref{fig:ex_density_map}.

By eye, the images in Figure \ref{fig:ex_density_map} that were obtained using the WGAN clearly resemble those that were obtained using N-body simulations.  To assess differences in the two sets of images that may be present at a level that is too low to be detected by eye, we compute two statistics: [1] the distribution of the normalized matter density contrast in the individual image pixels and [2] the matter power spectrum.

\begin{figure}
    \centering
    \includegraphics[
    width=\columnwidth]{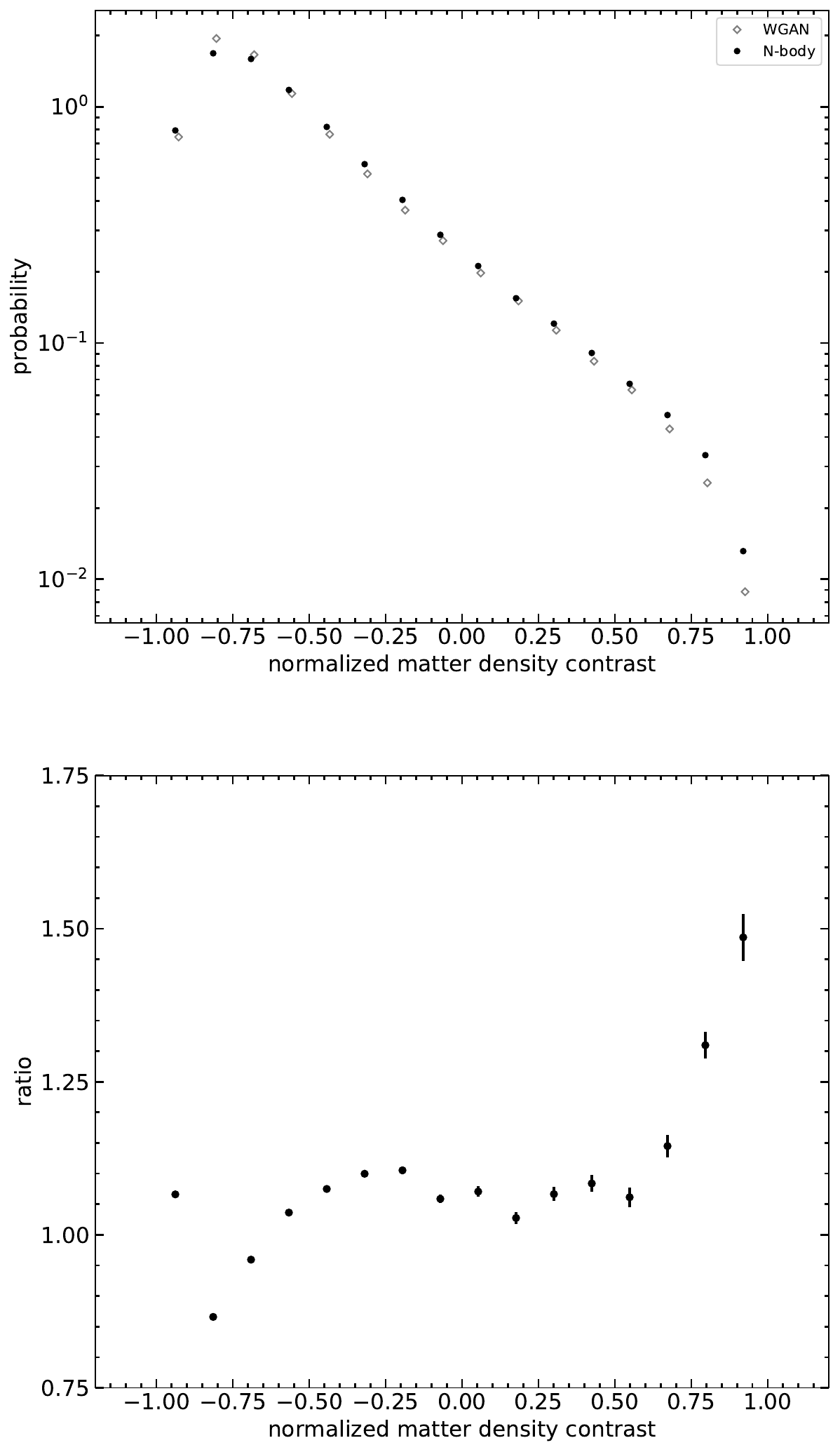}
    \caption{\textbf{Top:} Normalized probability distributions for the matter density contrast in the individual pixels of the images. Diamonds: results from 15,000 images generated with a WGAN.  Circles: results from 15,000 simulated images.  Error bars are omitted when they are comparable to or smaller than the sizes of the data points. \textbf{Bottom:} Ratio between the probability distributions shown in the top panel.}
    \label{fig:density_hist}
\end{figure}

The top panel of Figure \ref{fig:density_hist} shows the normalized probability distribution for the matter density contrast in the individual pixels.  All generated and simulated images were used to construct the probability distribution.
The bottom panel of Figure \ref{fig:density_hist} shows the ratio of the probabilities in the top panel.
Error bars were computed using using 10,000 bootstrap resamplings of the data and are omitted when they are comparable to or smaller than the data points.

From Figure \ref{fig:density_hist}, we can see that, compared to the N-body results, the WGAN significantly under produces pixels that have the highest density contrast.  These pixels are, however, relatively rare occurrences in the images (i.e., they have low probability) and they do not occur within voids.  Potentially more problematical for void statistics are discrepancies between the frequencies at which pixels with the lowest density contrasts (i.e., the most underdense pixels)
occur.  These underdense pixels are the most common occurrences in the images (i.e., they have high probability), and differences in the distribution of density contrasts within these pixels should lead to slightly different statistics for both the central density contrast and the mean interior density contrast when we compare voids in the N-body images to voids in the GAN-generated images.
Compared to the N-body results, the WGAN under produces pixels with the lowest density contrast (i.e., the likely void centers) and over produces pixels with density contrasts in the range
$\sim -0.88$ to $\sim 0.63$.  This will lead to systematic differences in the distributions of the mean central densities and mean interior densities of the voids in the simulated and generated images.

\begin{figure}
    \centering
    \includegraphics[width=\columnwidth]{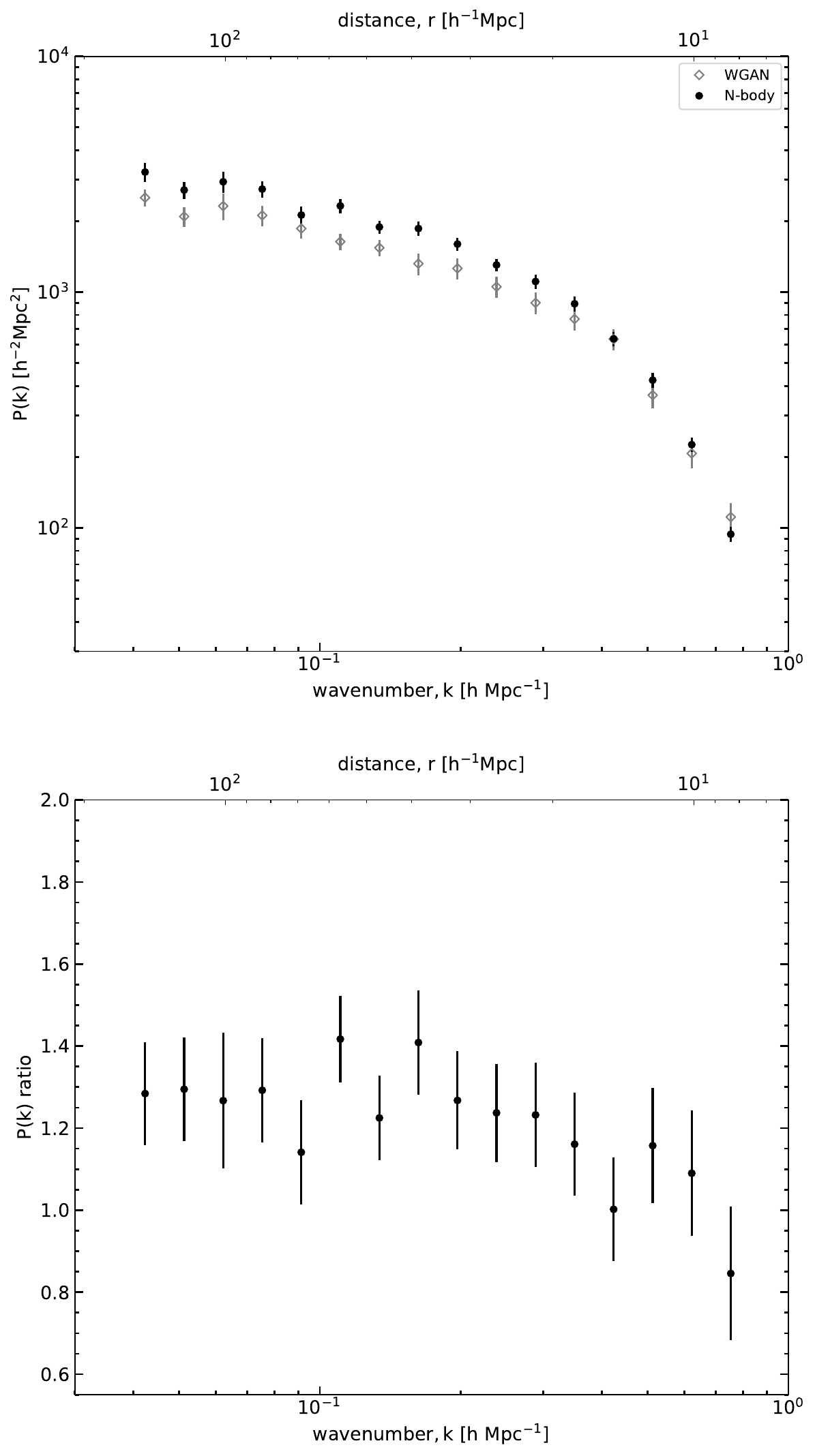}
    \caption{\bf{Top}: Mean matter power spectra, $P(k)$, computed from 15,000 simulated images (circles)
    and 15,000 images generated with a WGAN (diamonds). Error bars were computed using
    10,000 bootstrap resamplings of the data. \textbf{Bottom}: Ratio of the mean matter power spectra shown in the top panel.}
    \label{fig:pk}
\end{figure}

Figure \ref{fig:pk} shows the results for the matter power spectra of our simulated and generated datasets. 
The power spectra were calculated via 
\begin{equation} \label{eq:powerSpec}
    \langle\hat{\delta}(\vv{\bm{k}}) \hat{\delta}(\vv{\bm{k}}')\rangle = (2\pi)^{3}\delta\textsubscript{D}(\vv{\bm{k}}-\vv{\bm{k}}')P(\vv{\bm{k}})
\end{equation}
(see, e.g., \citealt{davis1985evolution})
and they encompass distance scales that are typical of void sizes (i.e., $10-100 h^{-1} \rm{Mpc})$.
Here, 
$\delta\textsubscript{D}(\vv{\bm{k}})$ is the Dirac delta function, $P(\vv{\bm{k}})$ is the 1D matter power spectrum, and $\hat{\delta}(\vv{\bm{k}})$ is the discrete Fourier transform of our density contrast field, defined as
\begin{equation} \label{eq:density_map}
    \delta = \frac{n}{\overline{n}} - 1 ~~~.
\end{equation}
Here $n$ is the number of particles in a cell and $\overline{n}$ is the mean number of particles in a field. The discrete Fourier transform was computed with Python's NumPy\footnote{\url{https://numpy.org/}} package \citep{numpy}.
Points in the top panel of Figure \ref{fig:pk} show the mean power spectra, computed using all simulated and generated images. Error bars were obtained from 10,000 bootstrap resamplings of the data. Points in the bottom panel of Figure \ref{fig:pk} show the ratio of the mean power spectra from the simulated images to the mean power spectra from the generated images.
From Figure \ref{fig:pk}, $P(k)$ from the generated images agrees with $P(k)$ from the simulated images on scales $< 20h^{-1} \rm{Mpc}$.  On larger scales, however,
the generated images yield systematically lower values of $P(k)$ than do the simulated images.

\subsection{Underdensity Probability Function} \label{sec:vpf}

In the observed universe, the Void Probability Function (VPF) is commonly used to determine
whether a randomly selected region in an image is devoid of galaxies (e.g., \citealt{whitevpf}; \citealt{Lachieze-Rey}; \citealt{vogleyvpf}). The VPF depends on the 
galaxy $N$-point correlation function via 
\begin{multline} \label{eq:vpf}
    P_0(N, A) = \\ exp\left[\sum_{N=1}^{\infty} \frac{(-n)^N}{N!} \int_{A}^{} w_N(\vv{\bm{x}}_1,...,\vv{\bm{x}}_N)d^2 x_1...d^2 x_N\right]
\end{multline}
(e.g., \citealt{whitevpf}),
where $n$ is the average number density of galaxies in the field, $w_N$ are the $N$-point correlation functions, and $\vv{\bf{x}}_i$ are the i-th galaxy positions in the area $A$. 

Since our GAN outputs consist of images of a smoothed mass density distribution (i.e., not 
the locations of individual galaxies or individual particles) we do not compute the VPF.
Instead, we compute a similar statistic known as the Underdensity Probability Function (UPF; see, e.g., \citealt{vogleyvpf} and \citealt{Tinker2006}). The UPF measures the probability that, within an area of radius $r$,
a randomly selected region of space is less dense than some particular threshold density. An advantage to using the UPF over the VPF is that the VPF is sensitive to shot noise because it requires counting individual particles or galaxies within a given aperture. 

To identify underdense regions in the density maps, we define a normalized density contrast $\Delta \equiv \frac{\delta\rho}{\rho}$ and we adopt a threshold for
the mass density contrast of $\Delta < -0.6$. Here $\rho$ is the local matter density and $\delta\rho$ is the difference between the local matter density and mean matter density of the image. That is, the UPF is defined as $U(r) = P(\Delta < \Delta_{\rm crit}; r)$ where $\Delta_{\rm crit} = -0.6$. The value of $\Delta_{\rm crit} = -0.6$ was chosen to match the mean void underdensities reported in \cite{Nico2014} and \cite{sanchez2016cosmic}.

The UPF for the simulated and generated images was determined by computing the mean mass density
contrast within randomly-placed circles of radius $r$.  If the average density contrast within
a circle of radius $r$ is less than $\Delta_{\rm crit}$, then that
particular region qualifies as being sufficiently underdense to be included in the calculation
of the UPF. A total of 
$N_{\rm test} = 10^4$ random circles of radius $r$ were placed within each image and the
total number of underdense regions, $N_0$, was then computed.  The UPF was then defined as
$U(r) = N_{(0;~ \Delta < \Delta_{\rm crit})} / N_{\rm test}$. 

\begin{figure}
    \centering
    \includegraphics[width=\columnwidth]{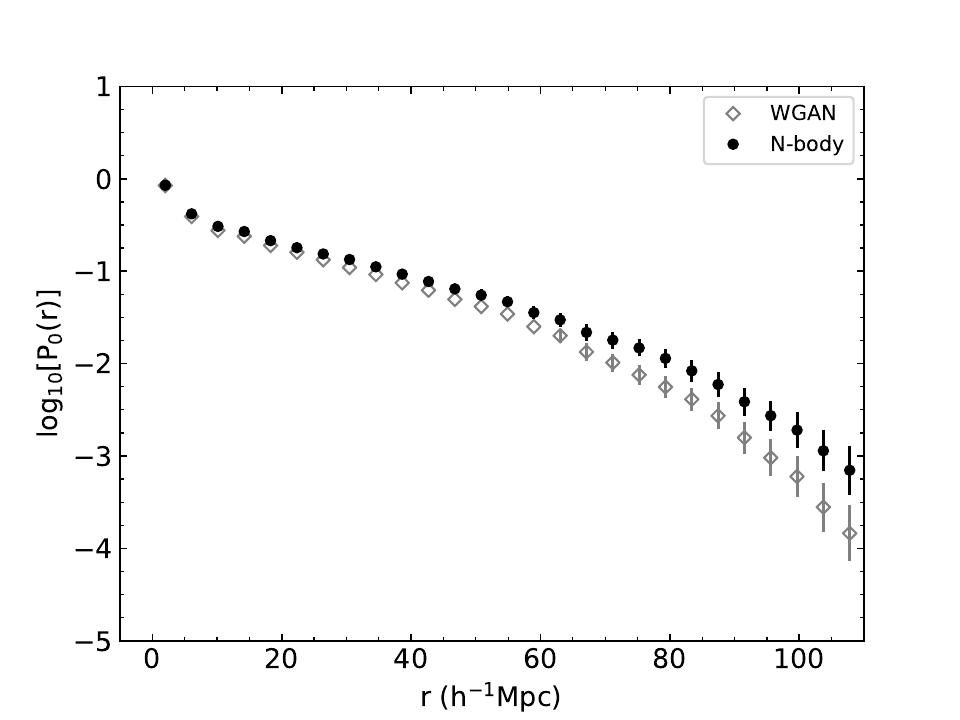}
    \caption{Mean underdensity probability function for images obtained from 
    the simulated images (circles) and the WGAN-generated images (diamonds). 
    Error bars were computed from 10,000 bootstrap resamplings of the 
    data and are omitted when they are comparable to or smaller than the data points.
   }
    \label{fig:UPF}
\end{figure}

Figure \ref{fig:UPF} shows the mean UPF obtained from
15,000 simulated images and 15,000 generated images. The UPFs were calculated following the steps described above and error bars were calculated by bootstrap resampling of both sets of distributions. Note that, since smaller circles in an image are encompassed by larger circles
in the same image, the data points
in Figure \ref{fig:UPF} are not independent of one another.  Hence, the point-to-point scatter is considerably smaller than the formal, bootstrapped error bars.
Similar to the power spectra, there is good agreement between between the simulated and WGAN images on small scales. On larger scales, there is a lower chance of finding underdense regions in the generated data than in the simulated data, with the difference increasing for regions $\geq 60 h^{-1}$~Mpc.  

\subsection{Void Identification} \label{sec:voidID}

To identify voids,
we adopted a \textit{voidfinder} algorithm that is similar to the 
algorithm used by \cite{voidfinder} and \cite{hoylevoidsin2dfs} 
to identify voids in catalogs of galaxies. 
When applied to an observational dataset, the \textit{voidfinder} algorithm 
identifies voids using the observed locations of galaxies.  In our case there
are no galaxies in the images.  Instead, the images consist of pixels with known
values of the mass density.  Our version of the \textit{voidfinder} algorithm,
therefore, scans images of the large-scale structure as defined by the pixellated
mass density.  The algorithm yields a set of individual void centers and a list
of which pixels in a given image belong to which voids.

Our \textit{voidfinder} algorithm can be summarized in four steps: [1] each pixel
is classified as being either a `wall pixel' or a `void pixel', [2] the distance
between every void pixel and the nearest wall pixel is calculated, [3] void
centers are defined to be the locations 
of the most isolated void pixels (i.e.,
the void pixels that are farthest from a wall pixel), and 
[4] as part of the process of identifying 
underdense regions that are interconnected, the areas of the larger voids are `enhanced' 
to include overlapping regions of low mass density. 
The first step is performed by scanning over every pixel in the image using
a circular aperture of radius $d=3$~pixels that takes the periodic boundary into account. A value of $d=3$ pixels (or $6h^{-1}$~Mpc) was chosen as a sensible size scale with which to judge a pixel's local environment.

If the pixel that the circular aperture is centered on contains at least $N=9$ pixels 
with a normalized density contrast $> -0.6$,
the pixel is classified as a `wall pixel'. That is, if $\sim d^2$ of the pixels in the neighborhood around the central pixel are sufficiently dense compared to the mean, the central pixel is classified as a wall pixel. 
Any pixel that is not classified as a wall pixel is classified as a `void pixel.'
The second step is to iterate over each void pixel and determine its distance to the nearest wall pixel. This is done by finding the circle of maximum radius that can be centered around each void pixel such that the circle contains no wall pixels. Following the terminology of \cite{hoylevoidsin2dfs}, we refer to these maximal circles as `holes'.

Once all of the pixels have been classified as being either a void pixel or a wall
pixel, the void centers are then identified. To do this, we begin by sorting all holes in a given image 
from largest to smallest in terms of their radii. 
We then define the center of the largest hole in the image
to be the center of the first void. Having found this first void center, we then iterate over all holes. If the next hole in the sorted list overlaps with any previously-identified
void by more than 10\% of its own area, 
that hole is not considered to be an individual void and, at this point, is temporarily ignored. A value of 10\% was chosen to insure only the largest hole in a local area is classified as a void, while still accounting for smaller voids that interconnect the larger voids.
If the next hole in the sorted list does not overlap with any previously-identified void (or if the overlap is $\le 10$\% of its own area), the hole is classified as a distinct void, with the center of the hole being the void center. 
This process continues until all holes with 
radius $> 3$~pixels have been considered.



In order to account for the fact that underdense regions of space are
interconnected, the final step in the process of identifying the voids is to consider the
degree of overlap between holes and voids.
The process is intended to examine the underdense regions near the edges of the holes and 
either assign those underdense regions to appropriate voids or reject them as being part of a void.
If a hole overlaps a single void by $> 50$\% of its own area (i.e., if the center of the hole is within the void's maximal hole), then we consider the 
hole to be part of the void. If a hole overlaps multiple voids by $> 50$\% of its
own area, we consider it to be part of the largest overlapping void. 
If a hole overlaps a void by 
$\le 10$\% of its own area, then the hole is identified as a distinct void, but the area of
overlap is assigned to the larger void.  If the overlap between a hole and a void is $> 10$\%
of the hole's area but $< 50$\%, the pixels in the hole are not assigned to any void.  By
this process, the total areas of the larger voids are enhanced relative to the areas of
the maximal circles that are centered on the void centers. 

\begin{figure*}
   \centering
    \includegraphics[scale=0.95]{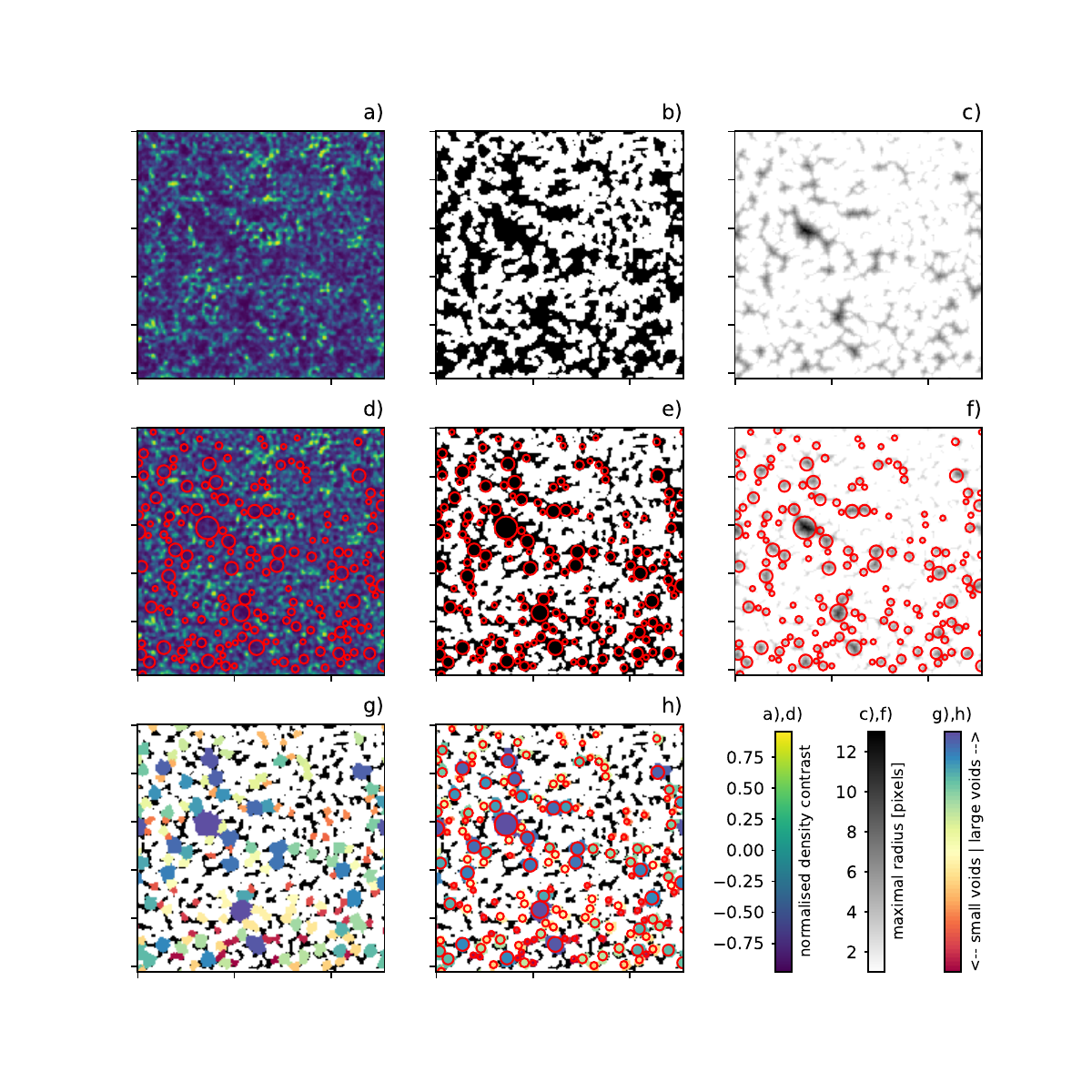}
    \caption{Illustration of the \textit{voidfinder} algorithm: a) normalized mass density, b) locations of wall pixels (white) and void pixels (black), c) maximal radius for each void pixel, d) -- f) largest hole contained within each void, overlaid on the panels
    directly above (red circles),  g) final void areas overlaid on the wall/void pixel distribution, h) same as (g), but with the largest hole contained within each void superposed on the image.
    }
    \label{fig:voidfinder}
\end{figure*}

Figure \ref{fig:voidfinder} shows the results of the \textit{voidfinder}
algorithm as applied to a single image. 
The different panels of Figure \ref{fig:voidfinder} show: a) the normalized density contrast, b) the locations of wall pixels (white) and void pixels (black), c) the distance to the nearest wall pixel at each location (i.e., the maximal radius for each pixel), d) the largest 
holes (i.e., the circles
that define the void centers) overlaid on the density contrast, 
e) the largest holes overlaid on the wall/void pixel image, f) the 
locations of the largest holes overlaid on the distance to the nearest wall pixel, g) 
the final void areas overlaid on the wall/void
pixel image, and h) same as g), but with the largest holes indicated.
The 15,000 images from the N-body simulations yielded a total of 2.5~million voids and the 15,000 generated images yielded a total of 2.8~million voids.  
The mean number
of voids contained within a single image from the N-body simulations ($169 \pm 10$) is, however, consistent with the mean number of voids contained within a single generated image ($184 \pm 11$).


\subsection{Void Properties} \label{sec:voidFinderResults}

Below we adopt the following terminology when discussing the void properties.  The `radius'
of a void is simply the radius of the void's maximal circle (i.e., the largest circle, centered on
the void center, with radius equal to the distance to the nearest wall pixel).  The
`effective radius' of a void corresponds to the radius of a circle that, after the area of the
void has been adjusted for any overlaps, is equal to the final area assigned to
the void.  Since some of the smaller voids lose area to neighboring large voids, their
effective radii will be less than the radii of their maximal circles.  Conversely, the
larger voids with areas that have been enhanced by the addition of pixels from neighboring
small voids (or significantly overlapping holes)
will have effective radii that are larger than the radius of their maximal
circle.

\begin{figure}
    \centering
    \includegraphics[ width=\columnwidth]{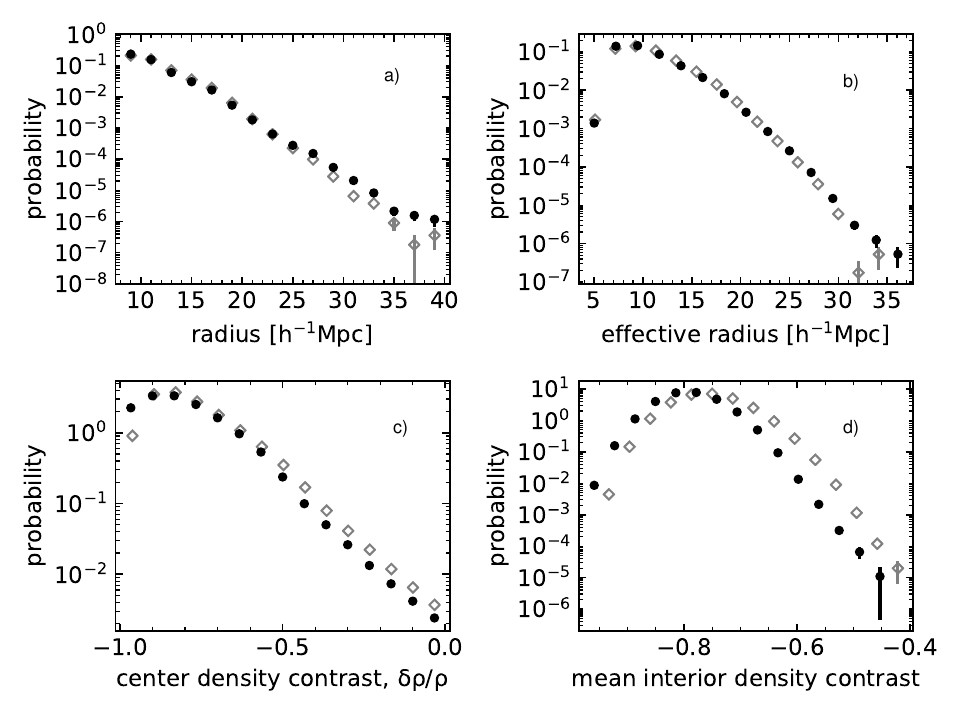}
    \caption{Normalized probability distributions for various void
    properties:  a) void radii,  b) void effective radii, c) value of the density contrast
    at the centers of the voids, d) mean interior density contrast, computed using the final
    areas of the voids.  Diamonds: results from images obtained with the WGAN.  Circles:
    results from images obtained from N-body simulations.
    Error bars are omitted when they are comparable to or smaller than
    the sizes of the data points.
    }
    \label{fig:r_re}
\end{figure}

The top panels of Figure \ref{fig:r_re} show normalized probability distributions for a) the void radii and b) the effective void radii.  Overall there is good agreement between the results for the simulated and generated images.  Compared to the simulated images, however, the generated images produce fewer of the smallest voids. The median void radius in the generated images is identical to
that of the simulated images ($10.0 h^{-1}$~Mpc), but the median effective void radius in the generated images ($10.0 h^{-1}$~Mpc) is somewhat larger than it is in the simulated images
($9.6 h^{-1}$~Mpc). 
Figure \ref{fig:r_re}c) shows the normalized probability distributions
for the density contrast at the centers of the voids. Here, again, the distributions are similar
and the median value of the central density contrast is nearly identical for the simulated
and generated images ($-0.83$ in the simulated images vs.\ $-0.81$ in the generated
images).  However, compared to the simulated images, the generated images 
produce far fewer voids in which the central density contrast is $\sim -1$ (i.e., the most 
underdense void centers).  Conversely, the generated images produce somewhat more
voids with central density contrasts between
$\sim -0.9$ and $\sim -0.5$ than do the simulated images.
Figure \ref{fig:r_re}d) shows the normalized probability distributions for the mean
density contrast within the voids. While the median values of the distributions in
Figure \ref{fig:r_re}d) are similar ($-0.79$ for the simulated images vs.\ $-0.76$
for the generated images), there is a clear offset between the two distributions
such that the mean density contrast of the generated voids is systematically higher than
that of the simulated voids. 

The results in Figure \ref{fig:r_re}c) and \ref{fig:r_re}d) are tied
directly to the results in Figure \ref{fig:density_hist}.  That is, compared to the
simulated images, the generated images contain systematically fewer of the least 
dense pixels.  Since voids are centered on the least dense pixels, this results in
fewer voids in the generated images with central density contrasts $\sim -1$.  It also results
in fewer voids in the generated images having mean interior densities $\lesssim -0.8$.
The combination of the generated images having both fewer of the least dense pixels
and more of the pixels with density contrasts in the range $\sim -0.88$ to $\sim -0.63$ gives rise to more voids with mean interior density contrasts $\gtrsim -0.7$ in the generated images

When averaged over all voids, the mean properties of the voids in the simulated and generated images are consistent within the formal error bars. In the simulated images, the mean radius and the mean effective radius are
$9.9 \pm 0.2 h^{-1}$~Mpc and $10.3 \pm 0.2 h^{-1}$~Mpc, respectively.  In the generated images, the mean radius and mean effective radius are
$10.1 \pm 0.2 h^{-1}$~Mpc and $10.5 \pm 0.2 h^{-1}$~Mpc, respectively.  In the simulated images, the mean central density contrast and the mean interior density contrast are $-0.81 \pm 0.01$ and $-0.79 \pm 0.01$, respectively.  In the generated images, the mean central density contrast and the mean interior density contrast are $-0.78 \pm 0.02$ and $-0.76 \pm 0.02$, respectively.  We note that, while the dispersions in the mean radius and the mean effective radius are identical for the voids in the simulated and generated images, the dispersions in the mean central density contrast and the mean interior density contrast are twice as large for the voids in the generated images as they are for the voids in the simulated images.

\subsection{Properties of the Largest Voids}

For the remainder of our analyses we will focus on a comparison of the best-resolved voids.  
These are the largest voids in the images, and here we
restrict the analyses to only those voids with radii $\ge 10$~pixels (corresponding to 
a radius $\ge 20h^{-1}$~Mpc).  Because large voids are rare, these
represent only a small fraction of the total number of
voids in the samples (15,063 in the simulated sample and 16,165 in the generated sample).  However,
unlike the smallest voids (which are poorly resolved), these largest voids are sufficiently
well resolved to allow a computation of their radial density profiles.

\begin{figure}
    \centering
    \includegraphics[width=\columnwidth]{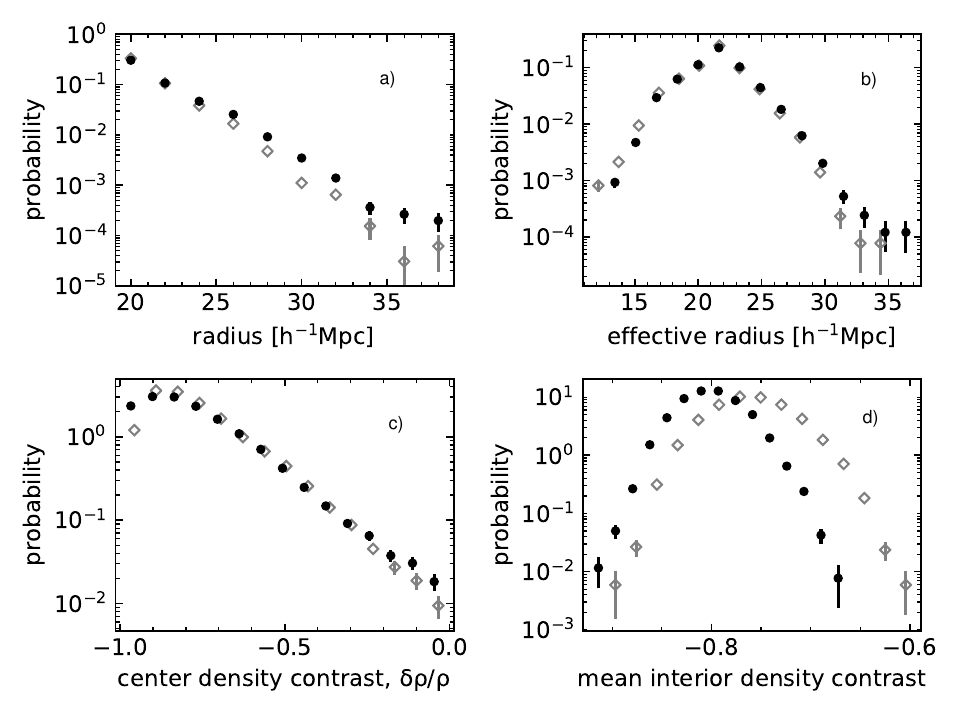}
    \caption{Same as Figure \ref{fig:r_re}, but here only voids with radius  
    $ \geq 20h^{-1}$ Mpc
    are used. 
    }
    \label{fig:large_voids_r_re}
\end{figure}

For comparison with our complete samples of voids shown in Figure \ref{fig:r_re}, 
Figure \ref{fig:large_voids_r_re} shows the same normalized probability distributions,
but using only the largest voids.  Compared to the results in Figure \ref{fig:r_re}, there
is a much greater difference between the distributions of void radii and void effective
radii in Figure \ref{fig:large_voids_r_re}.  This is due to a combination of two effects that
go into assigning each void its final area (from which the effective radius is determined).
First, the larger holes have a larger number of
underdense pixels just outside their circumferences than
do the smaller holes.  Therefore, a large hole that has been identified as being a void
has a higher probability of having its final area enhanced relative to the area of the 
hole (i.e., by the addition of pixels from smaller holes that overlap the large hole
by a significant amount).  This necessarily increases the effective radius of a void 
with an enhanced area.  

In addition, because of their relatively large size, the probability
of the largest voids in a given image overlapping 
one another by a small amount is higher than the probability of
the smallest voids in a given image overlapping by a similar amount.  In the case of these small
overlaps ($\le 10$\% of the area of the smallest of the two associated holes), the overlap
area is assigned to the larger void, resulting in the smaller of the two voids having
a final area that is reduced compared to the size of its hole (and hence, having an 
effective radius smaller than the radius of its associated hole).  From Figure \ref{fig:large_voids_r_re}a), the median values of the void radii 
($20.0 h^{-1}$~Mpc) are identical in the simulated and generated images.  Similarly,
the void effective radii ($21.4 h^{-1}$~Mpc) are identical in the simulated and generated images (see Figure \ref{fig:large_voids_r_re}b).  

Figure \ref{fig:large_voids_r_re}c) shows trends that are
similar to those in Figure \ref{fig:r_re}c).  While
the median central densities of the largest voids are similar in both the simulated 
and the generated images ($-0.82$ for the simulated images vs.\ $-0.81$ for the generated images),
the generated images produce far fewer of the largest voids with central densities
$\sim -1$.  Like the complete void sample, the generated images also produce somewhat
more large
voids with central densities greater than $\sim -1$, but the range over which this occurs
($\sim -0.9$ to $\sim -0.75$) is smaller than in the full sample.  Figure \ref{fig:large_voids_r_re}d)
also shows trends that are similar to those in Figure \ref{fig:r_re}d).  
The median values of the distributions in
Figure \ref{fig:large_voids_r_re}d) are similar ($-0.80$ for the simulated images vs.\ $-0.76$
for the generated images), and there is a clear offset between the two distributions
such that the mean density contrast of the largest generated voids is systematically higher than
that of the largest simulated voids.


\begin{figure}
    \centering
    \includegraphics[width=\columnwidth]{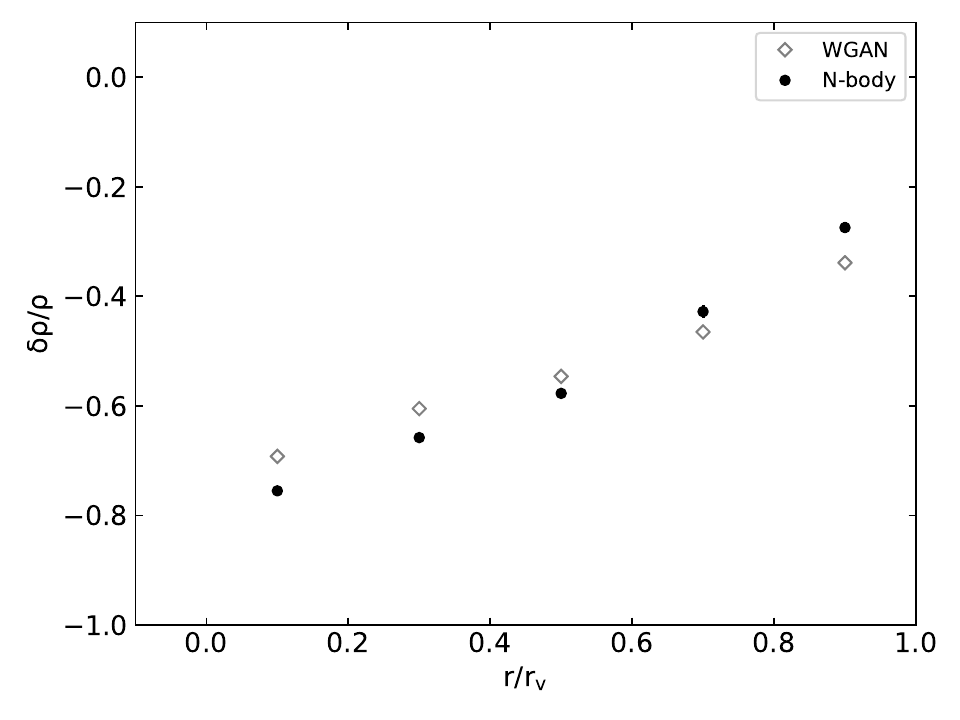}
    \caption{Mean radial underdensity profiles of largest voids, plotted as a function of 
    dimensionless distance relative to the void centers, ($r/r_v$).  Error bars
    are omitted because they are comparable to or smaller than the data points.
    }
    \label{fig:radial_profile}
    
\end{figure}

Lastly, Figure \ref{fig:radial_profile} shows the mean radial underdensity profiles
for the largest voids, plotted in terms of dimensionless distance relative to the void radius,
($r/r_v$).  As expected, the central regions of the
voids are extremely underdense and the density contrast increases with radius from 
the void centers.
While the radial underdensity profiles are similar in both the simulated and generated images, there
are clear differences.  For distances $r \lesssim 0.6 r_v$, the underdensities of the generated
voids are somewhat higher than the underdensities of the simulated voids.  For distances
$r \gtrsim 0.6 r_v$, the sense of the disagreement reverses, with the underdensities of
the generated voids being somewhat lower than the underdensities of the simulated voids.


\section{Discussion} \label{sec:discussion}

We used a standard deep convolutional
Wasserstein GAN (WGAN) to generate
novel 2D images of the smoothed mass density field of a $\Lambda$CDM
universe, and we explored the use of void statistics as a possible
method for assessing and fine-tuning GAN performance. 
The WGAN was trained using 15,000 2D images of the large-scale
structure that were obtained from N-body
simulations.  The trained WGAN was used to generate 15,000 2D images and a {\it voidfinder} algorithm, modified to identify regions of low mass density (as opposed to regions of
low galaxy density), was then used to create void catalogs from the simulated and
generated images.  The simulated and generated images yield a similar number of voids
(2.5 million voids and 2.8 million voids, respectively).

Within the formal error bars, the 
mean void properties (i.e., mean radius, mean effective radius, mean
central density contrast, and mean interior density contrast) in the simulated and
generated images are consistent with each other.
Systematic differences between the distributions of the void
properties are, however, apparent. Compared to the simulated images, the
generated images yield somewhat fewer of the smallest voids and significantly 
fewer voids with central density contrast $\sim -1$.  
Overall, the generated images yield fewer of the emptiest voids and, as a result, the
distribution of mean interior density contrast for the generated voids is offset
systematically from the distribution of mean interior density contrast for the 
simulated voids.

Neural networks function by learning the average trends in the target distribution 
but they struggle to capture absolute patterns in the data (see, e.g., \citealt{li2017diffusion} for a detailed discussion). 
That is, our deep convolutional WGAN is clearly able to detect trends in the pixel-to-pixel variations of the smoothed density contrast, 
but it is not reliably capturing the absolute highs and lows. The effect of this 
on void properties manifests in Figures \ref{fig:r_re}c) and
\ref{fig:large_voids_r_re}c), where the images yield significantly fewer voids 
with central density constrast $\sim -1$ than do the simulated images. 
The inability of the WGAN to reliably capture the absolute highs and lows of the 
smoothed density contrast also explains the systematic offsets between the distribution
of mean interior density contrast for the generated and simulated images
in Figures \ref{fig:r_re}d) and \ref{fig:large_voids_r_re}d).

GANs present an opportunity to investigate the distribution of the largest structures in Gpc-scale simulations
without the need for as many computationally expensive Gpc-scale N-body simulations as would otherwise be necessary.  In order for GANs to become a truly viable alternative 
to simulations, however, they need to be able to fully reproduce the structure that 
is seen in typical N-body simulations, including the frequency and properties
of voids. Because of this, void statistics are a natural higher-order test of the validity of GAN-generated 
maps of large-scale. 

While our WGAN produced voids with properties that
were in broad general agreement with the properties of voids in N-body simulations, the
systematic differences between the populations are an indication that improvements 
in the WGAN approach are still necessary.
For future work, the network architecture will need to be optimized in such a way that it
allows the trained WGAN to better capture absolute trends in the target images. Such
optimization may be dependent on the size of the
convolutional kernels and/or the values of the hyperparameters used 
in the generator and critic networks. 
The use of inception blocks (see \citealt{inception}), for example, would help 
the network to pick up features of varying sizes by scanning the input layer 
with kernels of varying sizes. 
This alone might resolve the issue of the centers of the generated voids 
being insuffciently underdense. 
Finally, it will be interesting to see whether
more sophisticated network architectures, such as the one used by \cite{kodi2020super} to generate
large-scale structure maps, yield improved results for the statistics of generated voids.

\section*{Acknowledgements}
We are grateful for insightful comments from two anonymous reviewers that helped to improve the paper.

\section*{Funding}

This research did not receive any specific grant from funding agencies in the public, commercial, or not-for-profit sectors.



\bibliography{references}

\appendix
\section{Wasserstein GAN} \label{sec:wgan appendix}

\cite{WGAN} suggest switching the standard GAN loss function to the Wasserstein metric (a.k.a.\ the Kantorovich–Rubinstein metric or Earth mover's distance). An intuitive explanation of the Earth mover's distance is found in its name.  Imagine trying to relocate a pile of dirt from one location to another. The Earth mover's distance is then the metric that provides the most cost effective way to transport the dirt. For WGANs, the loss function becomes:
\begin{equation} \label{eq:wgan}
    W(\mathbb{P}_{data}, \mathbb{P}_{gen}) := \inf_{\gamma(x,y) \in \Pi(\mathbb{P}_{data}, \mathbb{P}_{gen})} \mathbb{E}_{(x,y)\sim\gamma} [||x-y||] ~~.
\end{equation}
Here, $\Pi(\mathbb{P}_{data}, \mathbb{P}_{gen})$ is the set of all couplings of $\mathbb{P}_{data}$ and $\mathbb{P}_{gen}$. If each distribution is thought to be a unit amount of soil piled over a metric space, then $\gamma(x,y)$ represents how much soil must be moved from $x$ to $y$ such that $\mathbb{P}_{data}$ becomes $\mathbb{P}_{gen}$. In other words, Equation \ref{eq:wgan} reveals the cost of the optimal way to transfer soil.


In practice, solving Equation \ref{eq:wgan} is intractable.
However, \cite{WGAN} have shown that the Wasserstein metric reduces to 
\begin{equation}\label{eq:reduced_wgan}
    W(\mathbb{P}_{data}, \mathbb{P}_{gen}) := -\mathbb{E}_{\vv{\bm{z}}\sim p_{prior}(\vv{\bm{z}})}[f(G_{\theta_G}(\vv{\bm{z}}))] ~~,
\end{equation}
provided that $f$ exists and is K-Lipschitz. In the WGAN algorithm, $f$ is analogous to the set of weights, $w$, that parameterize the critic network. 
In other words, given weights, $w$, lying in a compact space, $W$, one back propagates through $\mathbb{E}_{\vv{\bm{z}}\sim p_{prior}(\vv{\bm{z}})}[f_{w}(G_{\theta_G}(\vv{\bm{z}}))]$ to update the generator's weights, just as one would do in the standard GAN architecture. However, to ensure the weights lie in $W$, the weights are clipped to a box $W=[-c,c]$, where $c$ is a clipping parameter.

The training algorithm for a WGAN is similar to that of the standard GAN algorithm. 
Aside from the different loss functions and weight clipping, the only change in the WGAN algorithm is that the critic network has its weights updated an integer times more than the does the generator network (i.e., in the standard GAN architecture, both $G$ and $D$ are updated evenly). This change is possible because, unlike the Jensen-Shannon Divergence, the
Earth mover's distance is continuous and differentiable everywhere.

Allowing the critic to train in a more stable manner 
makes mode collapse improbable in the WGAN architecture. This is due to the fact
that the critic is no longer trained on how well it can discriminate between
real and fake images.  Rather, the critic is trained on how it scores the
realness or fakeness of an image. Training the critic in this
way improves network stability by giving the generator better feedback on 
how to update its weights.

\end{document}